\documentclass[%
superscriptaddress,
preprint,
 amsmath,amssymb,
]{revtex4-2}

	\usepackage{bm}
	\usepackage{braket}
	\usepackage{dsfont}

 	\usepackage{graphicx}
	\usepackage[caption=false]{subfig}
	\usepackage[export]{adjustbox}
	\usepackage{tabularx}

	\usepackage{xcolor,colortbl}
\usepackage[english]{babel}
\usepackage{amsmath}
\usepackage{amssymb, comment}
\usepackage[]{graphicx}
\usepackage{times}
\usepackage{bm}
\usepackage{units}
\usepackage{braket}
\usepackage[version=3]{mhchem}

\usepackage{bbm}
\usepackage{tabularx}

\usepackage{color}

	\newcommand{\vect}[1]{\boldsymbol{#1}}		
	\newcommand{\op}[1]{\hat{\boldsymbol{#1}}}	
	

\begin{document}

\title{Phonon-bottleneck enhanced exciton emission in 2D perovskites}

\author{Joshua J. P. Thompson}
\email{joshua.thompson@physik.uni-marburg.de
}
\affiliation{Department of Physics, Philipps-Universit\"{a}t Marburg, Renthof 7, 35032 Marburg}
\affiliation{Department of Materials Science and Metallurgy, University of Cambridge, Cambridge CB3 0FS, United Kingdom}
\author{Mateusz Dyksik}
\affiliation{Department of Experimental Physics, Faculty of Fundamental Problems of Technology, Wroclaw University of Science and Technology, 50-370 Wroclaw, Poland}
\author{Paulina Peksa}
\affiliation{Department of Experimental Physics, Faculty of Fundamental Problems of Technology, Wroclaw University of Science and Technology, 50-370 Wroclaw, Poland}
\affiliation{Laboratoire National des Champs Magnetiques Intenses, 143 Avenue de Rangueil 31400 Toulouse, France}
\author{Katarzyna Posmyk}
\affiliation{Department of Experimental Physics, Faculty of Fundamental Problems of Technology, Wroclaw University of Science and Technology, 50-370 Wroclaw, Poland}
\affiliation{Laboratoire National des Champs Magnetiques Intenses, 143 Avenue de Rangueil 31400 Toulouse, France}
\author{Ambj\"{o}rn Joki}
\affiliation{Department of Physics, Chalmers University of Technology, Gothenburg 412 96, Sweden}
\author{Raul Perea-Causin}
\affiliation{Department of Physics, Chalmers University of Technology, Gothenburg 412 96, Sweden}
\author{Paul Erhart}
\affiliation{Department of Physics, Chalmers University of Technology, Gothenburg 412 96, Sweden}
\author{Michał Baranowski}
\affiliation{Department of Experimental Physics, Faculty of Fundamental Problems of Technology, Wroclaw University of Science and Technology, 50-370 Wroclaw, Poland}
\author{Maria Antonietta Loi}
\affiliation{Zernike Institute for Advanced Materials, University of Groningen, Nijenborgh 4, 9747 AG Groningen, The Netherlands}
\author{Paulina Plochocka}
\affiliation{Department of Experimental Physics, Faculty of Fundamental Problems of Technology, Wroclaw University of Science and Technology, 50-370 Wroclaw, Poland}
\affiliation{Laboratoire National des Champs Magnetiques Intenses, 143 Avenue de Rangueil 31400 Toulouse, France}
\author{Ermin Malic}
\affiliation{Department of Physics, Philipps-Universit\"{a}t Marburg, Renthof 7, 35032 Marburg}

\date{\today}

\begin{abstract}
    Layered halide perovskites exhibit remarkable optoelectronic properties and technological promise, driven by strongly bound excitons. The interplay of spin-orbit and exchange coupling creates a rich excitonic landscape, determining their optical signatures and exciton dynamics. Despite the dark excitonic ground state, surprisingly efficient emission from higher-energy bright states has puzzled the scientific community, sparking debates on relaxation mechanisms. Combining low-temperature magneto-optical measurements with sophisticated many-particle theory, we elucidate the origin of the bright exciton emission in perovskites by tracking the thermalization of dark and bright excitons under a magnetic field. We clearly attribute the unexpectedly high emission to a pronounced phonon-bottleneck effect, considerably slowing down the relaxation towards the energetically lowest dark states. We demonstrate that this bottleneck can be tuned by manipulating the bright-dark energy splitting and optical phonon energies, offering valuable insights and strategies for controlling exciton emission in layered perovskite materials that is crucial for optoelectronics applications.
\end{abstract}

\maketitle
\onecolumngrid

Hybrid metal-halide perovskites represent a unique material system \cite{schilcher2021significance, egger2018remains}, bridging the electronic properties of epitaxial and organic semiconductors. They emerged as revolutionary materials for photovoltaics \cite{snaith2019decade}, before a plethora of other possible applications were proposed, from light emission \cite{tan2014bright} to quantum optical technologies\cite{tamarat2020dark, cai2023zero}, where perovskite nanostructures show particular promise \cite{yuan2017one, tsai2021bright, gong2018electron}.  Both nanocrystals and 2D layered perovskites are known for their superior emissive properties \cite{gong2018electron,tamarat2019ground, becker2018bright}.   Their quantum emission efficiency is one or two orders of magnitude higher than in epitaxial inorganic semiconductors, however the origin of this technologically important characteristic is still the subject of ongoing research and debate\cite{tamarat2019ground,becker2018bright, baranowski2020excitons}.

The optical response of perovskites is governed by excitons exhibiting a characteristic fine structure comprising bright triplet and dark singlet states \cite{becker2018bright,tanaka2005electronic, fu2017neutral, tamarat2019ground}. Recent studies have shown that the dark state is situated several to tens of meV below the bright states \cite{tamarat2019ground,tamarat2020dark,dyksik2021brightening,posmyk2022quantification}. Despite this significant splitting, perovskites exhibit surprisingly intense PL emission even at cryogenic temperatures \cite{tamarat2019ground,becker2018bright,gong2018electron, kutkan2023impact} making them very attractive for quantum technology. Moreover, in nanocrystals, intense emission is simultaneously observed for all bright states even though they are also separated by several meV \cite{fu2017neutral,tamarat2019ground}. This emissive behaviour indicates that the exciton population in perovskites does not follow Boltzmann statistics, however, the origin of this high bright-state occupation has remained elusive. Thus, an in-depth quantum mechanical modeling of exciton relaxation dynamics including the interplay of the exciton fine structure and scattering with phonons is required to fully exploit and design perovskites for light emission applications. 

Here, we address this problem for an archetypical perovskite, (PEA)$_2$PbI$_4$, where PEA stands for phenylethylammonium. This 2D layered perovskite demonstrates significant splitting between the bright and dark states ($\sim$20meV) \cite{dyksik2021brightening,posmyk2022quantification} which is off-resonant with optical phonon modes\cite{straus2016direct,urban2020revealing, thouin2019phonon, menahem2021strongly, ziegler2020fast}, indicating a possible bottleneck effect. Using a microscopic and material-specific many-particle theory, we explore the formation, relaxation and decay dynamics of excitons in this structure and show that the energy mismatch between the fine structure of exciton and phonons leads to a pronounced phonon-bottleneck effect. The consequence is an inefficient exciton relaxation to the energetically lowest dark state resulting in an enhanced non-thermal population of bright excitons, explaining the surprisingly high cryogenic PL emission. We find an excellent agreement between  theoretical predictions and  temperature-dependent magneto-optical spectroscopy measurements offering access to the dark exciton population.  Our work provides a comprehensive microscopic picture of the phonon-bottleneck effect in the family of 2D perovskites, explaining why these materials exhibit strong emission at low temperatures despite their dark ground state.

\section{Results}

\begin{figure}[t!]
    \centering
    \includegraphics[width=0.6\linewidth]{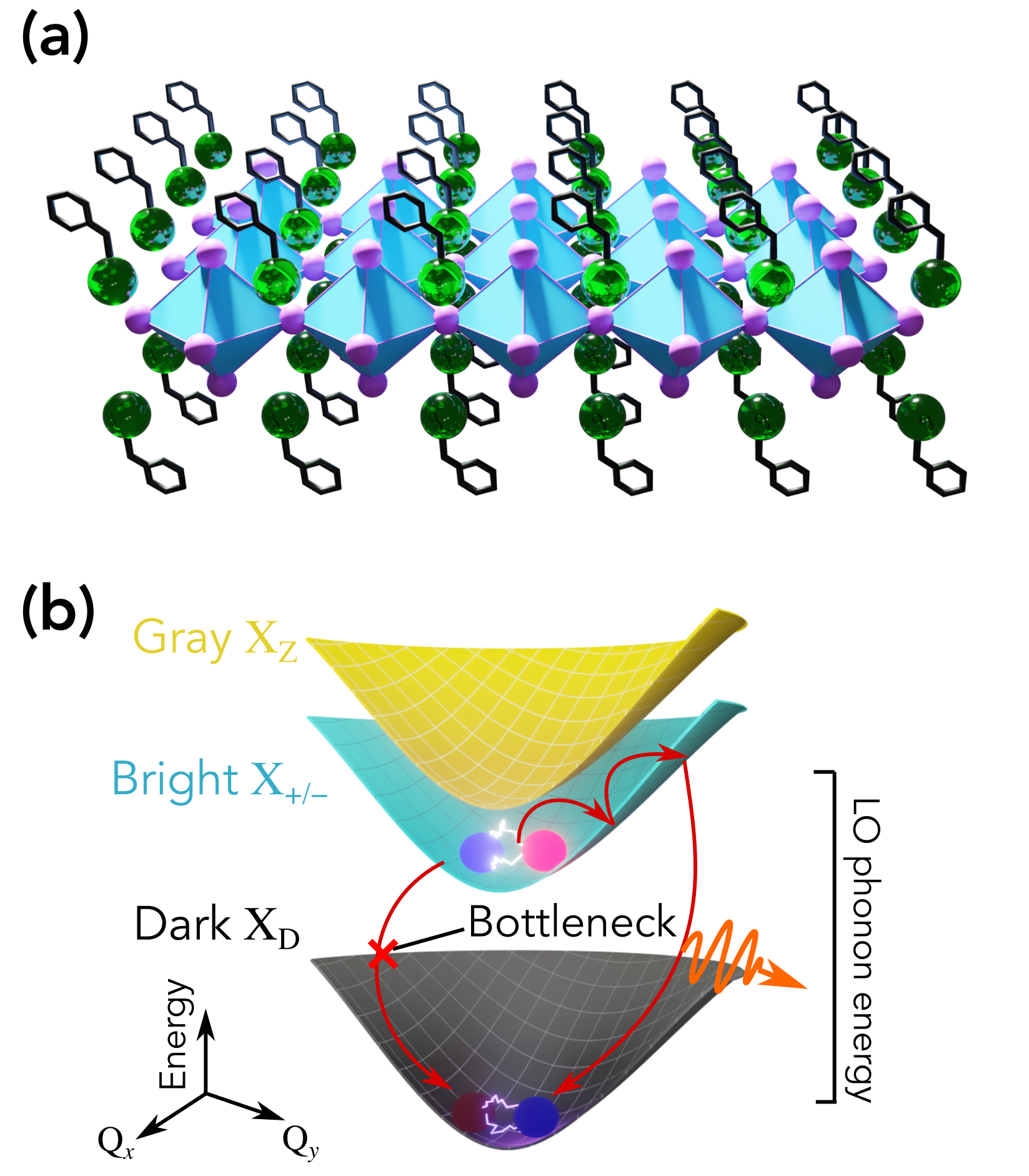}
    \caption{(a) Crystal structure of the 2D perovskite (PEA)$_2$ PbI$_4$.  (b) Schematic of the exciton fine structure including bright ($X_{+}$, $X_{-}$), gray ($X_\text{Z}$) and dark exciton states ($X_\text{D}$) and their relaxation dynamics. In particular,  the phonon-bottleneck is shown to hinder the scattering between bright and dark excitons.   }
    \label{fig1}
\end{figure}

\textbf{Exciton energy landscape:} Strong spin-orbit coupling and exchange interaction in perovskite nanocrystals and 2D perovskites give rise to an exciton fine-structure comprised of a dark singlet and bright triplet states \cite{fu2017neutral, becker2018bright, sercel2019exciton, tanaka2005electronic, ghribi2021dielectric, quarti2023exciton}. Recent experimental studies have provided extensive characterization of exciton states as well as their energy spacing in (PEA)$_2$PbI$_4$\cite{posmyk2022quantification, dyksik2021brightening, posmyk2023exciton, do2020bright}. In layered perovskites (cf. Fig.\,\ref{fig1}(a)) the lowest energy dark state $X_D$ is typically separated by $\sim 15-20$\,meV from the two in-plane bright states $X_{+}$ and $X_{-}$ and by another $\sim$1 meV from the out-of-plane polarised grey state $X_{Z}$. A schematic of the exciton state structure is presented in Fig.\,\ref{fig1}(b). We discuss the impact of different crystal phases in the SI.
  
We model the electron-hole interaction with the generalised Keldysh potential \cite{kira2006many} (all parameters can be found in SI). The energy landscape including the material-specific splitting has been determined microscopically by solving the Wannier equation\cite{feldstein2020microscopic, koch2006semiconductor} before including the impact of the short- and long-range exchange interaction. Our model accurately recovers qualitatively and quantitatively all characteristic features of the exciton fine structure \cite{dyksik2021brightening, posmyk2023exciton, posmyk2022quantification, biffi2023excitons, quarti2023exciton} observed in the exemplary (PEA)$_2$PbI$_4$, a reliable basis for studying the excitonic relaxation dynamics.

\begin{figure}
    \centering
    \includegraphics[width=0.6\linewidth]{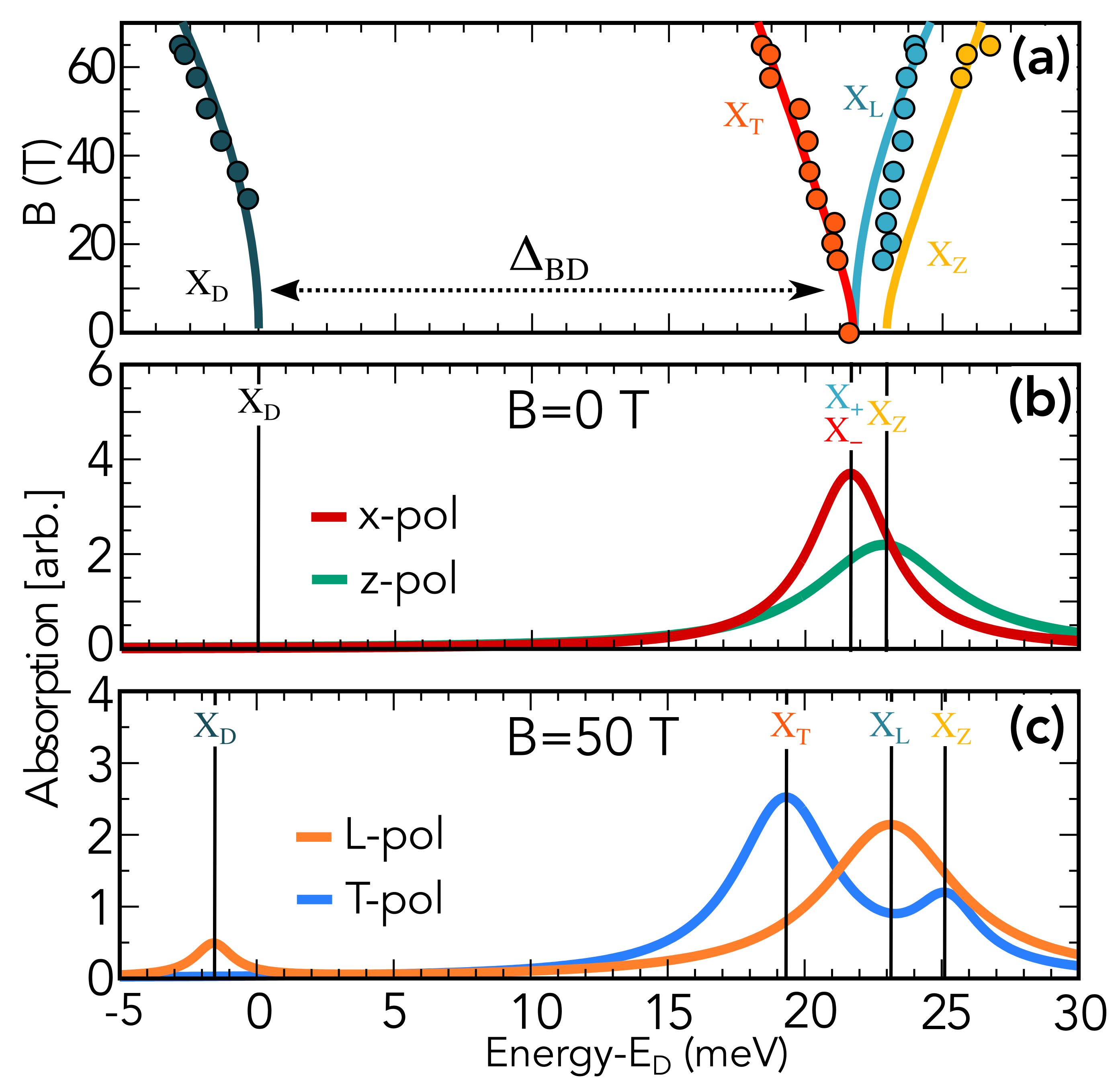}
    \caption{(a) Exciton fine-structure of the 2D perovskite (PEA)$_2$ PbI$_4$ as a function of in-plane magnetic field, calculated (lines) and measured (dots). The bright-dark splitting, $\Delta_\text{BD}$, is shown with the arrow. As the magnetic field increases, the polarisation of  longitudinal ($X_L$) and dark ($X_D$) as well as the polarisation of  gray ($X_Z$) and  transverse polarised ($X_T$) excitons mix. (b), (c) Absorption spectra at 30 K for $B=0$ T and $B=50$ T, respectively. The positions of exciton states are marked with thin vertical lines.  Note that without a magnetic field, there are ciruclary polarised states ($X_{\pm}$) that become linearly polarised ($X_\text{L,T}$) due to mixing at $B \neq 0$. }
    \label{fig2}
\end{figure}

A quantitative picture of the theoretically predicted (lines) and measured (dots) excitonic landscape can be seen in Fig\,\ref{fig2}\,(a), which shows the spectral shift of the dark, bright and grey states as a function of the applied in-plane magnetic field (Voigt configuration). At zero field, we find a bright-dark splitting, $\Delta_\text{BD}$, of 21\,meV. 
The magnetic field lifts the degeneracy of the bright in-plane states $X_{+/-}$ and changes their polarization from circularly to linearly polarised. The new in-plane states are now polarised parallel (longitudinal $X_L$) and perpendicular (transverse $X_T$) to the magnetic field. We perform magnetic field and polarisation-dependent transmission measurements to experimentally determine the excitonic energy landscape. We find an excellent theory-experiment agreement for the magnetic field dependence of the exciton energy, cf.\,Fig.\,\ref{fig2}\,(a).

The magnetic field mixes the dark $X_D$ and the gray $X_Z$ states with the superposition of the two bright in-plane states \cite{surrente2021perspective}, resulting in a partial transfer of the oscillator strength from $X_L$ and $X_T$ to $X_D$ and $X_Z$ states, respectively. To investigate this, we determine the optical absorption spectra (at 30\,K) of (PEA)$_2$PbI$_4$ by numerically evaluating the Elliot formula and second-order Born-Markov equation \cite{kira2006many, feldstein2020microscopic, brem2018exciton} (see SI for more details). In the absence of a magnetic field (B=0\,T), only absorption related to the bright triplet states can be observed: two degenerate circularly polarized states or a grey state, cf.  Fig. \ref{fig2}(b). For x-polarised absorption (red line in Fig. \ref{fig2}(b)), signatures from the $X_{+/-}$ excitons are observed, while the gray exciton cannot be seen. On the other hand, for z-polarised light (green line) the gray state is visible, while the bright states are not. The enhanced broadening of the z-polarised emission comes from additional phonon scattering channels to the lower-lying bright excitons (cf. Fig. \ref{fig1}(b)). 
The lowest energy spin-dark exciton $X_D$ has a vanishing oscillator strength and cannot be seen, however it can be optically accessed in a magnetic field, cf. Fig.\,\ref{fig2}(c). Here, the dark state becomes visible and L- and T-polarised bright excitons are now clearly distinct. Due to its efficient mixing with the $X_L$ state, the dark exciton can be only observed in the L-polarised absorption (orange line), whereas the gray exciton is only visible for T-polarised excitation (blue line) reflecting its mixing with the $X_T$ state.

\textbf{Phonon-bottleneck effect:} Following optical excitation of the bright states, excitons thermalise via scattering within or between excitonic bands. Considering a clean sample in the low-excitation regime, exciton relaxation is driven by phonon-induced scattering. For most semiconductors, fast phonon-mediated scattering leads to a thermalized Boltzmann distribution \cite{brem2018exciton}. However, in systems where phonon-assisted scattering is prohibited due to inherent differences between the initial and final states (mismatch of angular momentum, energy or momentum), a non-thermal distribution can arise. 
In 2D perovskites, strong spin-orbit coupling in the conduction band mixes different states, facilitating  scattering between excitons with opposite spin. Another possible origin for a non-thermal distribution is a mismatch between the energy of the initial and final states and the involved phonon energies. In the case of (PEA)$_2$PbI$_4$, it was observed that the excitonic transition is predominantly coupled to the optical phonon with the energy of $\sim$ 35\,meV \cite{zhang2018optical, urban2020revealing, ziegler2020fast, feldstein2020microscopic, straus2016direct} that is significantly larger than the bright-dark splitting $\Delta_\text{BD}\sim20$ meV (c.f. Fig.\ref{fig2}(a)). 

As a consequence, to scatter from the bright to the dark states, hot excitons need to be formed with considerable excess energy of at least $15$ meV to compensate the energy difference between the bright-dark energy splitting $\Delta_\text{BD}$ and the phonon energy $E_\text{LO}$. Therefore, optically excited excitons in the light-cone first need to scatter up to higher-momenta states by absorption of acoustic phonons, cf. Fig. \ref{fig1}b. This process is highly unlikely at cryogenic temperatures resulting in a phonon-bottleneck effect, i.e. exciton cannot scatter to the energetically lowest states and thus cannot build a thermal Boltzmann distribution. Note that while Fig. \ref{fig1}b shows a typical initial exciton occupation centred at small momenta, the final exciton distribution is independent of the initial excitation conditions, as excitons get trapped when they reach the minimum of the bright exciton band.

\begin{figure}
    \centering
    \includegraphics[width=0.75\linewidth]{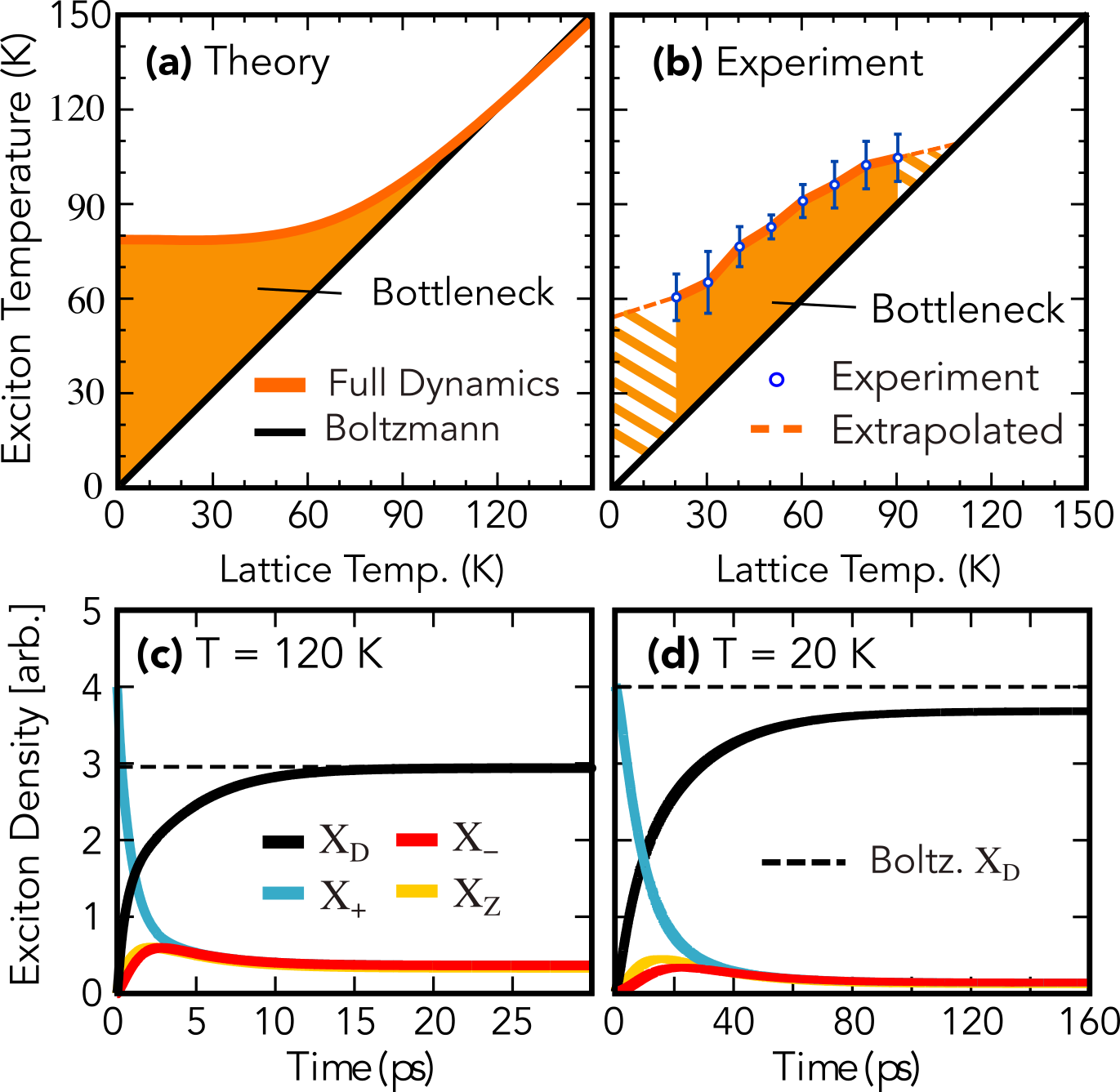}
    \caption{ (a) Effective steady-state temperature $T_\text{exc}$ of the bright exciton as a function of the lattice temperature T. The shaded region indicates the phonon-bottleneck effect, i.e. $T_\text{exc}$ higher than expected from a Boltzman distribution (black line). (b) Experimentally obtained exciton temperatures (blue dots) with a linear extrapolation at low and higher temperatures, where a measurement was not possible (dashed orange). (c), (d) Theoretical calculations of the temporal evolution of the exciton density at 120 K and 20 K, respectively, following a resonant excitation of the bright exciton without a magnetic field. Dashed horizontal line in (c) and (d) shows the expected population of dark exciton assuming a Boltzmann distribution.  }
    \label{fig3}
\end{figure}

To corroborate this explanation, we derive a coupled set of semiconductor Bloch equations describing the temporal evolution of the momentum-resolved exciton occupation $\op{N}^{\vect{Q}}_n(t)$  based on the Heisenberg equation of motion \cite{brem2018exciton} (see Methods and SI for  details). Here, $n$  and $\vect{Q}$ denote the exciton state and its centre-of-mass momentum, respectively. This allows us to track the phonon-mediated relaxation cascade within the excitonic fine structure resolved in time and momentum.  A thermal equilibrium is obtained when the population of excitonic states becomes time-independent. To quantify the final distribution of excitons across the exchange-split states in 2D perovskites, we define an effective excitonic temperature, $T_\text{exc}$, as obtained by rearranging the Boltzmann distribution
$
    T_\text{exc} = \frac{\Delta_\text{BD}}{k_B} \left[\ln\left(\frac{N_D}{N_{+}}\right)\right]^{-1}.
$
The excitonic temperature is determined by the ratio of $N_D$ and $N_{+}$ corresponding to the exciton density of the dark and the bright exciton $X_D$ and $X_{+}$, respectively. Note that $N_{+}$ and $N_{-}$ are degenerate and thus equally populated at thermal equilibrium without a magnetic field.

By definition, in the case of a Boltzmann distribution, the exciton temperature is equal to the lattice temperature (black line in Fig. \ref{fig3}a). If excitons are trapped in the higher-energy state, $X_{+}$, then the effective exciton temperature will be larger than the lattice temperature reflecting the excess energy of hot excitons. To determine the excitonic temperature,  we assume an initial excitation of the  $X_{+}$ bright state achieved by applying a circularly polarised laser pulse. We then calculate the time- and momentum-resolved population dynamics of the four excitonic states by evaluating the semiconductor Bloch equations until the system is thermalised.
In Fig. \ref{fig3}a, the effective steady-state temperature of the bright exciton in the (PEA)$_2$PbI$_4$ is shown as a function of the lattice temperature $T$.  For $T >110 $K, we find that the effective exciton temperature $T_\text{exc}$ (orange line) coincides with the lattice temperature as determined by the Boltzmann distribution (black line). At lower temperatures, however, we show that $T_\text{exc}$ clearly deviates from the lattice temperature, approaching a constant value of around 75\,K even as $T\rightarrow 0 K$. This signature is a clear indication of a phonon-bottleneck effect, as excitons cannot relax to the energetically lower state and thus exhibit an excess energy that makes them hotter than the lattice temperature. At higher temperatures, this bottleneck effect can be circumvented by the absorption of acoustic phonons lifting bright excitons to higher states and allowing them to scatter down to the dark state, cf. Fig. \ref{fig1}b.

We directly compare our theoretical prediction to the experimentally obtained effective exciton temperature in Fig.\ref{fig3}b. The latter is obtained by magneto-optical PL measurements considering the relative oscillator strength of bright and dark exciton in the presence of a magnetic field (see SI for details). The experimental results show a clear deviation from the expected thermal Boltzmann distribution for temperatures lower than 100 K - in excellent agreement with the presented calculation. As the lattice temperature approaches 90 K, we observe that the phonon-bottleneck effect vanishes and the effective exciton temperature goes towards the lattice temperature. This clearly supports the message of a thermally activated phonon-assisted relaxation circumventing the phonon-bottleneck at higher temperatures.

To understand the origin of the phonon-bottleneck, we resolve the population dynamics of different exciton species in Fig.\ref{fig3}c,d at 120 and 20 K, respectively. The initial population of the optically excited $X_{+}$ exciton (blue line) begins to decay populating the dark ($X_D$) and gray ($X_Z$) states.  There is no direct coupling between the bright states, as this transition would require the conduction and valence band spin to flip simultaneously \cite{becker2018bright}, which is prohibited due to the absence of spin-orbit coupling in the valence band \cite{becker2018bright, kiss2016elliott}. Instead, excitons from the $X_{+}$ state scatter first into the dark or the grey state, before the $X_{-}$ state can be populated. This explains the delayed onset of the $X_{-}$ population. Since the bright and grey state are almost energetically equivalent, their final populations are very similar. The dark state is much lower in energy and therefore is the most populated at any temperature. At $T=120$ K, (Fig. \ref{fig3}(c)), the relative steady-state population follows Boltzmann statistics, as seen from the intersection between the dashed Boltzmann line and the dark exciton population after the equilibrium has been reached. 
In Fig. \ref{fig3}d., the 15 meV required to enable interband scattering between bright and dark states cannot easily be satisfied at 20 K, where the thermal energy $k_B T\approx 6.5$ meV is too low. Therefore, the relaxation process slows down significantly and the quasi-thermal equilibrium is only reached after tens of ps. Note that exciton lifetimes in layered perovskites ($\sim$ 40  ps \cite{fang2020band}) are  longer than the typical thermalisation time even at 20\,K.  At such a low temperature,  the entire excitonic population would effectively reside within the dark exciton band under Boltzmann statistics, as indicated by the dashed line in Fig. \ref{fig3}d. However, in our dynamics simulations, we find the dark exciton population to be significantly lower and hence the population of optically active excitonic states is orders of magnitude larger than expected from a Boltzmann distribution.  

\begin{figure}
    \centering
    \includegraphics[width=0.6\linewidth]{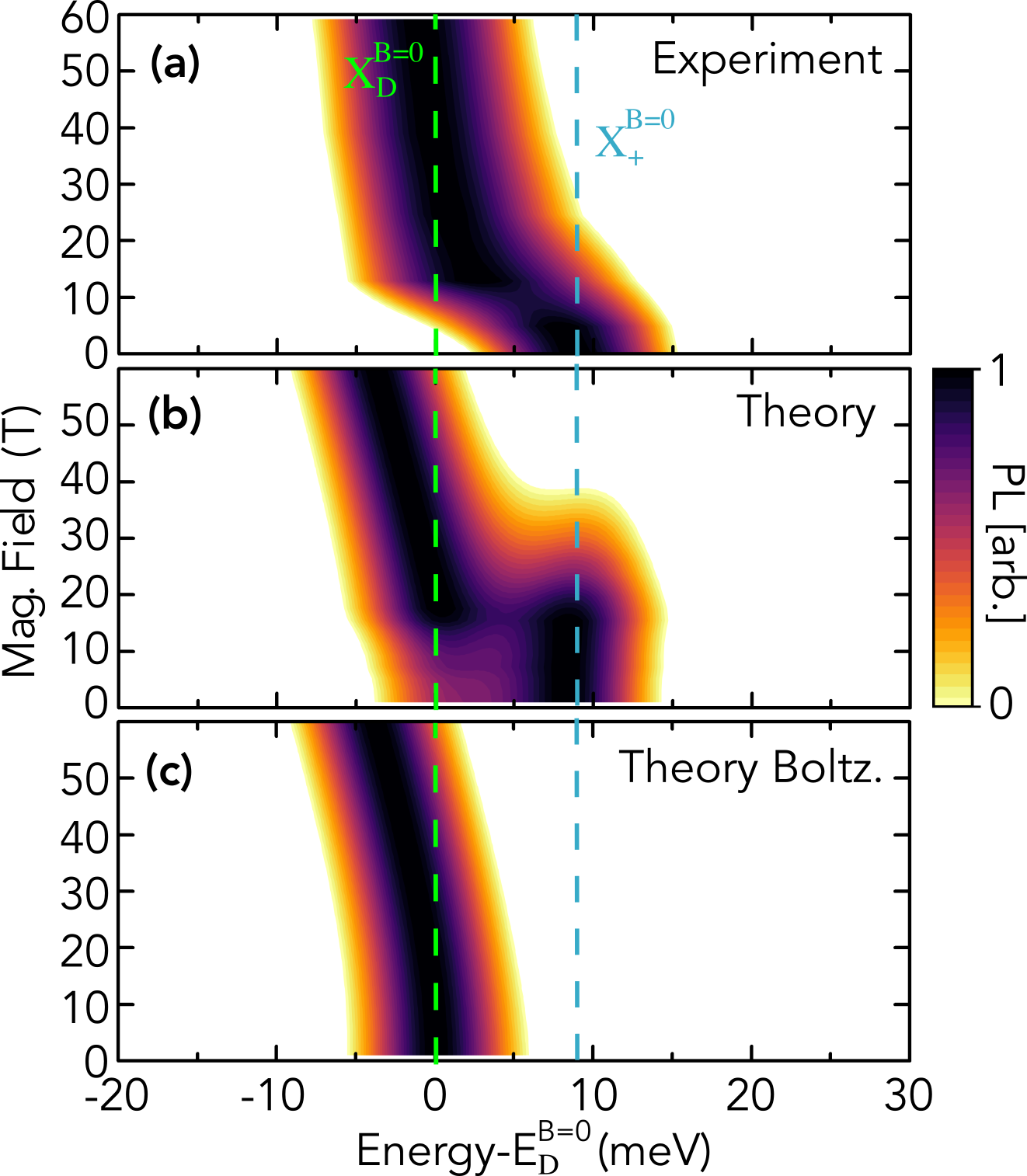}
    \caption{ (a) Normalised experimental PL spectra as a function of magnetic field at T= 40 K.  Theoretically predicted steady-state PL considering the full exciton dynamics (b) and assuming a Boltzmann distribution (c).  Dominant exciton resonances at zero magnetic field are indicated with dashed vertical lines.}
    \label{fig4}
\end{figure}

To further strengthen the proposed picture of the pronounced phonon bottleneck effect in (PEA)$_2$PbI$_4$, we show the experimental PL spectrum as a function of a magnetic field at 20 K - in direct comparison to our theoretical prediction after a stationary distribution has been reached. The magnetic field shifts the excitonic energy levels, but the quantitative effect is relatively small ($\sim$ 1-2 meV), and hence has little effect on the relaxation.  Therefore, we do not observe a significant effect of the magnetic field on the phonon-bottleneck. We distinguish in theory two cases: taking into account the full exciton dynamics (Fig. \ref{fig4}b) and assuming a Boltzmann distribution (Fig. \ref{fig4}c).  We fix the bright-dark splitting to about 9 meV as extracted from the PL spectra. The reduced splitting compared to the value found in the absorption spectra \cite{dyksik2021brightening} (Fig. \ref{fig2}c) is attributed to the Stokes shift, stemming from disorder red-shifting the bright exciton, and furthermore to a crystal distortion changing the crystal bond lengths and altering the exchange coupling that is responsible for the bright-dark energy splitting \cite{dyksik2021brightening}.
In the experiment, we observe a transition of the dominant PL signal from the bright to the dark state at about $B =$ 15 T.
Note that a non-zero oscillator strength of the dark state is observed in the experimental PL at $B=0$ T (Fig. \ref{fig4}a and Fig.\,S3), which is attributed to crystal distortion and higher-order optical processes \cite{dyksik2021brightening, tanaka2005electronic, quarti2023exciton, shinde2023emissive}. As such the dark state is always slightly visible in experiments. Therefore, we add a non-zero optical matrix element for the dark state also in our simulations.

 As the magnetic field is increased, the mixing between the bright and dark states becomes more efficient, such that previously dark exciton gains a progressively larger bright exciton component even dominating the PL spectrum above a certain critical value of the magnetic field. 
In theory, this occurs at larger values considering the full dynamics (Fig. \ref{fig4}b), whereas the dark state dominates already at zero field when assuming a Boltzmann distribution (Fig. \ref{fig4}c). This is caused by the much larger population of the energetically lowest dark exciton provided that there is no phonon bottleneck. The situation is different when considering the full dynamics, where excitons are trapped in the bright state and cannot scatter further down to the dark exciton, cf. Fig. \ref{fig3}. The interplay between the relative population of bright and dark excitons and the brightening of the dark exciton in the presence of a magnetic field determines which exciton dominates the PL.  
The theoretical prediction of a transition from the bright to the dark-exciton-dominated PL agrees well with the experiment and is yet further evidence of a phonon-bottleneck effect. Considering a fully thermalized Boltzmann distribution, there is no such transition, as shown in Fig. \ref{fig4}c.  The quantitative discrepancy between experiment and theory for the critical magnetic field value can be in part attributed to the higher effective exciton temperature predicted by the theory. The latter slightly overestimates the bright exciton population  possibly due to small, temperature-dependent changes in the phonon dispersion or excitonic structure not accounted for in our model.

 \begin{figure}[t!]
        \includegraphics[width=0.6\linewidth]{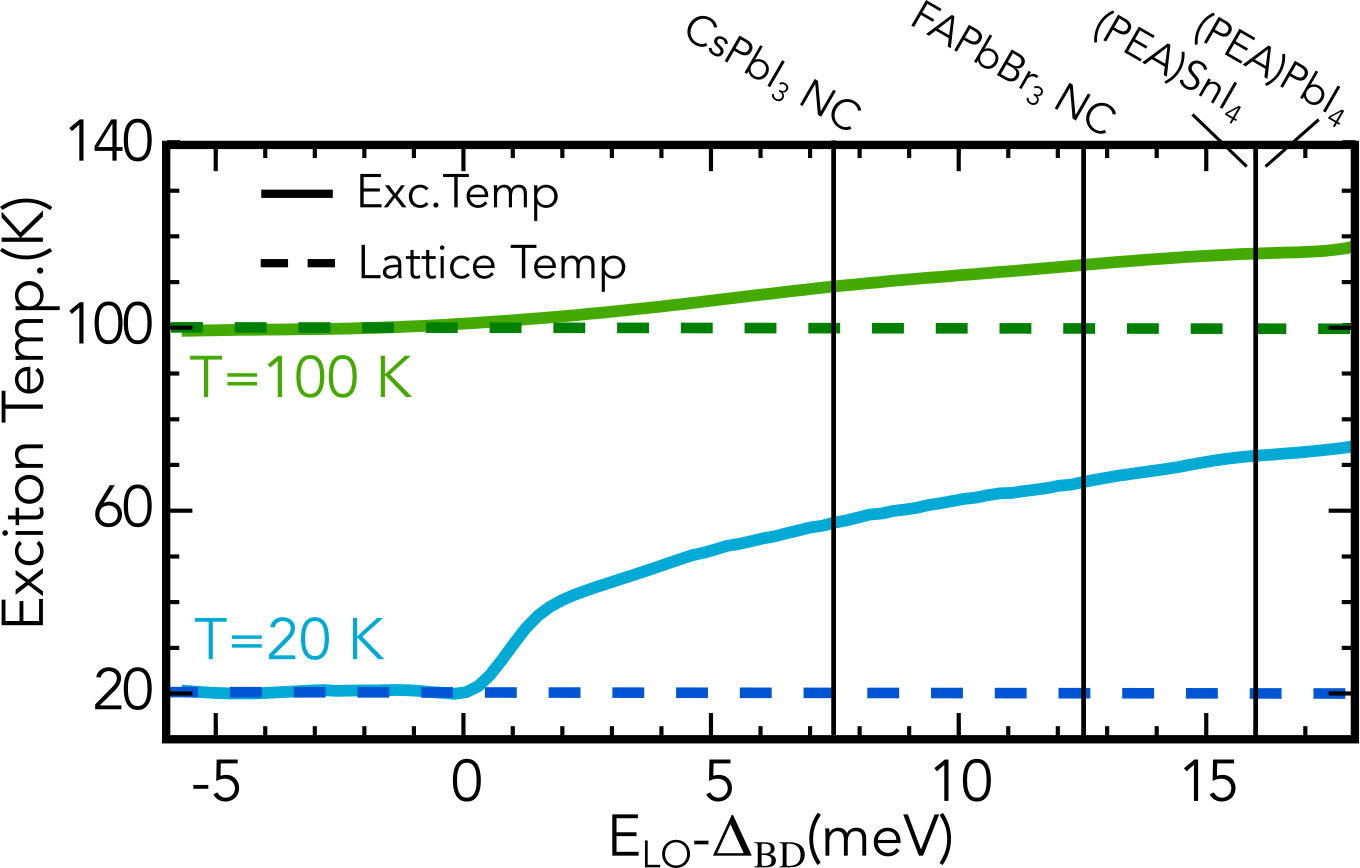}
    \caption{Demonstration of a raised effective exciton temperature as a function of the offset between the  optical phonon energy ($E_\text{LO}$) and the dark-bright energy splitting ($\Delta_\text{BD}$). For a vanishing offset, the exciton temperature matches the lattice temperature (dashed line). Examples of offsets measured in other materials are marked with vertical lines including CsPbI$_3$ nanocrystal (NC) \cite{fu2018unraveling, tamarat2020dark}, FAPbBr$_3$ NC \cite{tamarat2019ground}, (PEA)SnI$_4$ \cite{dyksik2021brightening} and (PEA)PbI$_4$ (this work).}
    \label{fig5}
\end{figure}
 
For a given perovskite, the difference between the bright-dark energy splitting $\Delta_\text{BD}$ and the optical phonon energy determines whether and at which temperature the bottleneck occurs. If there exists a phonon mode which couples to these excitons with an energy smaller than or resonant to $\Delta_\text{BD}$, the bottleneck will not occur.  
In Fig. \ref{fig5}, we show the effective exciton temperature as a function of the energy difference between the dominant longitudinal optical mode ($E_\text{LO}$) and the dark-bright exciton splitting $\Delta_\text{BD}$ for 100 K (green line) and 20 K (blue line). We find that for $(E_\text{LO}-\Delta_\text{BD}) \rightarrow 0$ the exciton temperature converges to the lattice temperature, as expected from a Boltzmann distribution. In contrast, as this energy difference increases, the effective exciton temperature grows larger demonstrating the existence of a phonon-bottleneck. We find the onset of this bottleneck to shift to lower $\Delta_\text{BD}$ the smaller the temperature is, as it becomes more difficult to overcome the energy offset with the thermal energy.

\section{Discussion}
 We have explored exciton optics and dynamics in 2D perovskites, combining a sophisticated microscopic theory with cryogenic magneto-optical spectroscopy measurements. We determined the exciton fine-structure  and the brightening of dark excitons in magneto-PL measurements. We tracked the relaxation dynamics of excitons and demonstrated in theory and experiment how a mismatch between the phonon energy and the splitting between bright and dark states gives rise to a pronounced phonon-bottleneck. This explains the observed unexpectedly strong emission of layered perovskites at low temperatures despite the dark ground state.  The phonon-mode dictating the phonon-bottleneck originates from the organic spacer layer \cite{straus2022photophysics}. Therefore, we predict that chemical engineering of the organic spacer \cite{boeije2023tailoring} can close or open the bottleneck. 
Overall, our joint theory-experiment study sheds new light on the population dynamics and optical response of layered perovskites,  crucial for optimizing the performance of perovskite-based devices.

\section{Methods}
\textbf{Exciton fine-structure:}
To model the optics and dynamics of excitons in layered perovskites, we first evaluate the Wannier equation \cite{kira2006many} to derive the exciton binding energies and wavefunctions, taking into account screening from the lead-halide layer itself and from the organic spacer layers. More details can be found in the supplementary information, including the form and parametrisation \cite{PhysRevB.45.6961, feldstein2020microscopic} of this interaction. We restrict our study to 1s excitons, since the 1s-2s separation is very large ($\approx 180$ meV) compared to the exciton fine-structure including dark, bright and gray states. The 1s binding energy is determined to be 230 meV, \cite{PhysRevB.45.6961, straus2016direct}, which is in good agreement with previous studies. While we assume a monolayer perovskite, stacking additional layers has little effect on the band structure, owing to the separation introduced by the organic spacer layer \cite{zhao2018layer}. The reduction in the excitonic binding energy from monolayer to bulk stems from the increased screening and is on the order of 10-20\% \cite{zhang2018optical}, which is much smaller than the monolayer-bulk comparison in other layered semiconductors, such as TMDs \cite{chernikov2014exciton} exhibiting a reduction of up to 90\%. Our results are therefore valid for a larger number of perovskite layers, with increased screening reducing the exchange interaction, further enhancing the mismatch with the optical phonon energy and enhancing the bottleneck.

The exciton fine-structure is determined by both the spin-orbit coupling and the exchange interaction. In layered perovskites,  the spin orbit coupling leads to a mixing of the orbital composition of the lowest conduction band \cite{becker2018bright}, which becomes a linear combination of $p$-orbitals.
In contrast, there is no mixing of the valence band $s$ orbitals. 
We derive an expression for the exchange interaction in the excitonic basis, taking into account the spin-orbit coupling in the conduction band. Details and the derivation can be found in the supplementary information.

\textbf{Exciton optics:}
The absorption spectra of the perovskite monolayer can be calculated using the Elliot formula \cite{kira2006many} where the absorption is described by a series of Lorentzian peaks centred at the exciton resonances  which takes into account both the radiative and non-radiative broadening. The absorption is calculated for $\sigma$-polarised light. The magnitude of the absorption signal is primarily determined by the optical matrix element. By taking into account the exchange interaction, the optical matrix elements and hence optical polarisation of the excitonic states can be determined.
The time-resolved photoluminescence can also be described using an Elliot formula, only now it is also proportional to the time-resolved occupation of the excitonic states, as detailed further in the SI.

\textbf{Phonon-driven exciton dynamics:}
We derive a series of semiconductor Bloch equations to capture the time evolution of the excitonic populations \cite{kira2006many, brem2018exciton}, following optical excitation of the bright state $X_{+}$.   We employ the Heisenberg equation of motion, $i\hbar\partial_t  N^n_{\vect{Q}} = \left\langle[N^n_{\vect{Q}}, \op{H}_X]\right\rangle$ to derive these equations. We apply the second-order Born-Markov approximation \cite{kira2006many} to truncate the equations to the most important terms, taking into account the exciton-phonon interaction.  We consider both the emission and absorption of optical and acoustic phonons, allowing excitons to thermalise both in momentum and between excitonic bands. The optical phonons are particularly crucial in driving the exciton relaxation, owing to the strict momentum and energy conservation for the scattering processes.   A detailed description is provided in the SI.
Due to the strong spin-orbit coupling in these materials, the conduction band spin can be flipped in an exciton-phonon scattering event.  For 2D perovskites, this process is necessary to enable phonon-scattering between the bright and dark/gray excitons.

\textbf{Optical spectroscopy:} The photoluminescence was excited with a continuous wave laser emitting at 407\,nm, in a nitrogen-cooled pulsed magnet, providing a maximum field of 60\,T with a pulse duration of 500 ms. The sample was installed in a variable temperature cryostat in the centre of the magnetic field. The measurements were performed in the Voigt configuration, with the $\mathbf{c}$-axis of the sample perpendicular to the magnetic field and parallel to the $\mathbf{k}$ vector of the exciting laser. Linear polarization was resolved in situ using a broadband polarizer.

\textbf{Sample fabrication:} Glass substrates were ultrasonically cleaned sequentially using detergent solution, deionized water, acetone, and isopropanol. Subsequently, the substrates were dried in an oven T = 140°C for $\ge$10 min before treatment with ultraviolet ozone for 20 min. Immediately after cleaning, the substrates were placed in a nitrogen-filled glove box for film deposition.
A stoichiometric precursor solution was used, prepared by dissolving PEA (Phenethylammonium iodide, 98.0\% TCI) and PbI2 at a molar ratio of 2:1 in a mixed solvent of N,N'- dimethylformamide and dimethyl sulfoxide (4:1 volume ratio, 0.5 M concentration). To homogenize the solutions, they were stirred for at least 3 hours at room temperature before deposition. A spin-coating process with antisolvent treatment was used to deposit the precursor solution onto the cleaned substrates. A rotation speed of 2000 rpm was used for the first 10 s of the spin-coating process. The speed was then accelerated to 8000 rpm for the remaining 30 s. Five seconds before the end of the spin-coating cycle, the antisolvent (chlorobenzene) was added to the substrate. The films were immediately annealed at 100°C in a nitrogen atmosphere for 10 min.

 \textbf{Acknowledgements: }
We acknowledge funding from the Deutsche Forschungsgemeinschaft (DFG) via SFB 1083 and the regular project 504846924. M.D. acknowledges support from the National Science Centre Poland within the SONATA grant (2021/43/D/ST3/01444). M.B. acknowledges support from the National Science Centre Poland  2021/43/I/ST3/01357.

\providecommand{\latin}[1]{#1}
\makeatletter
\providecommand{\doi}
  {\begingroup\let\do\@makeother\dospecials
  \catcode`\{=1 \catcode`\}=2 \doi@aux}
\providecommand{\doi@aux}[1]{\endgroup\texttt{#1}}
\makeatother
\providecommand*\mcitethebibliography{\thebibliography}
\csname @ifundefined\endcsname{endmcitethebibliography}  {\let\endmcitethebibliography\endthebibliography}{}

\end{document}